\documentclass[a4paper, 11pt]{article}
\usepackage[T1]{fontenc}
\usepackage[utf8]{inputenc}
\usepackage[english]{babel}
\usepackage{amsmath}
\usepackage{amssymb}
\usepackage{graphicx}
\usepackage{enumerate}
\usepackage{dsfont,color,units,pstricks}
\usepackage{amsfonts}

\newcommand{\beq}{\begin{equation}}
\newcommand{\eeq}{\end{equation}}
\newcommand{\be}{\begin{equation}}
\newcommand{\ee}{\end{equation}}
\newcommand{\beqa}{\begin{eqnarray}}
\newcommand{\eeqa}{\end{eqnarray}}

\newcommand{\nn}{\nonumber}

\begin{document}
\title{\Large Optimal Trading with Linear and (small) Non-Linear Costs}
\author{A. Rej$^1$, R. Benichou$^1$, J. de Lataillade$^1$, \\
G. Z\'erah$^1$ \& J.-Ph. Bouchaud$^{1,2}$\\
1: Capital Fund Management, \\ 
23 rue de l'Universit\'e, 75007 Paris, France \\
2: CFM-Imperial Institute of Mathematical Finance \\
Imperial College, London, SW7 2AZ, UK
}

\maketitle

\begin{abstract}
We reconsider the problem of optimal trading in the presence of linear and quadratic (market impact) costs for arbitrary linear costs but in the limit where quadratic costs are small. Using matched 
asymptotic expansion techniques, we find that the trading speed vanishes inside a band that is narrower in the presence of market impact by an amount that scales as a cube root of the market impact parameter. Outside the band we find three regimes: a small boundary layer where the velocity vanishes linearly with the distance to the band, an intermediate region 
where the velocity behaves as a square-root and an asymptotic region where it becomes linear again. Our solution is consistent with available numerical results. We determine the 
conditions under which our expansion is useful in practical applications and generalize our solution to other forms of non-linear costs.
\end{abstract}

\section{Introduction}

Determining the optimal trading strategy in the presence of a predictive signal and transaction costs is of utmost importance for quantitative asset managers, since too much trading (both in volume and frequency) 
can quickly deteriorate the performance of a strategy, or even make the strategy a money-losing machine. The detailed structure of these costs is actually quite complex. Some costs are called ``linear'' because they 
simply grow as $\Gamma Q$, where $Q$ is the traded volume and $\Gamma$ the linear cost parameter. These are due to various fees (market fees, brokerage fees, etc.) or the bid-ask spread and they usually represent a small fraction of the amount traded (typically 
$10^{-4}$ on liquid markets, but sometimes much more in OTC/illiquid markets).  Much more subtle are impact-induced costs, which come from the fact that a large order must be split into a sequence of small trades that are executed gradually.  But since each executed trade, on average, impacts the price in the direction of the trade, the average execution price is higher (if one buys) than the decision price, leading to what is called ``execution shortfall''.  This cost clearly increases faster 
than $Q$, since the price impact itself increases with the size of the trade. There seems to be a wide consensus now that the impact-induced costs are on the order of $\sigma Q^{3/2}/V^{1/2}$, where $\sigma$ is the daily
volatility and $V$ the daily turnover (see \cite{sqrt1,sqrt2,sqrt3} for recent accounts). 

\begin{figure}
\includegraphics[scale=0.5]{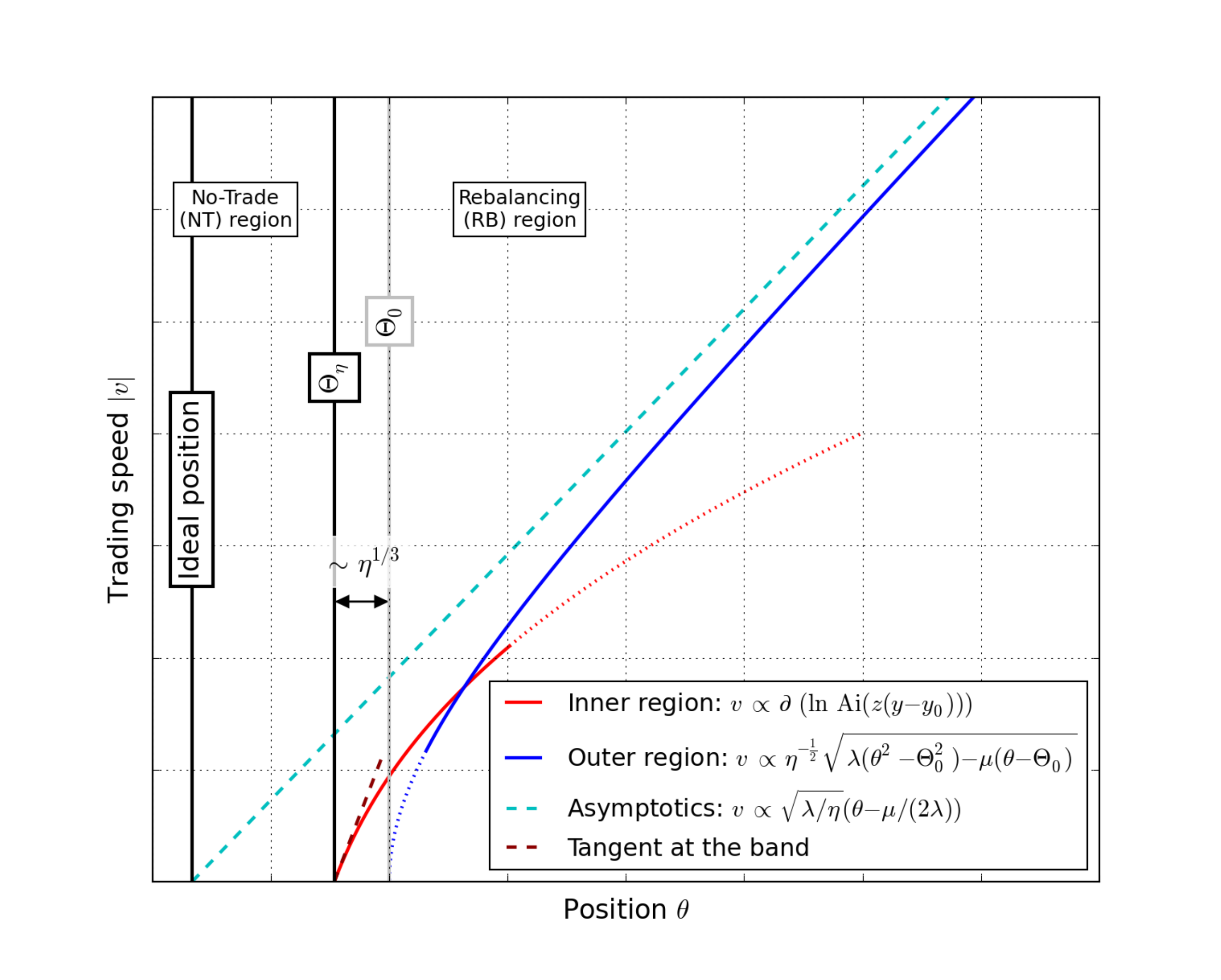}
\caption{Trading speed as a function of the distance to the boundary of the NT region. The outer solution behaves as a square-root close to 
the band and linearly farther out. The square-root singularity is regularized in a boundary layer of width $\eta^{1/3}$ so that the
velocity vanishes linearly at the (shifted) boundary of the band $\Theta_\eta$. } \label{fig:solution}
\end{figure}

From a theoretical point of view, however, the $Q^{3/2}$ dependence of the costs makes the analysis difficult. As a simplifying assumption, one often replaces the empirical $Q^{3/2}$ behaviour by a ``quadratic cost'' 
formula $\eta Q^2$, so that the price impact is proportionally to $Q$, see e.g. \cite{Almgren,OW}. In the absence of linear costs ($\Gamma = 0$), the optimal strategy may be found as a result 
of a simple quadratic optimisation problem, see for example \cite{Almgren,GarPed}. The optimal policy is to rebalance at finite speed towards the target portfolio. This results in a 
position that is an exponential moving average of the trading signal. The pure linear cost problem (i.e. $\eta=0$) was independently solved, in slightly different contexts, in \cite{MarSch, JCMJP}. It requires \textit{instantaneous} rebalancing towards a finite band around the ideal position, and no action inside the band, also called the no-trade (NT) region. The case where both linear and quadratic costs are present is 
of course highly interesting and no exact solution is known at this stage. An approximate solution was proposed in \cite{Sam}. A method for constructing the exact solution in the small cost limit where {\it both} $\Gamma$ and $\eta$ tend to zero can be found in \cite{LiuKarWeb}. 
The aim of our paper is to show that one can in fact relax the assumption that $\Gamma$ is small and expand around the general solution for linear cost, the expansion parameter being $\eta \to 0$. We will see that the solution defines four different regions (\textit{c.f.}  Figure \ref{fig:solution}): 
\begin{itemize}
\item a) the no-trade (NT) region inside a band around the ideal position is still present but the band shrinks by an amount $\sim \eta^{1/3}$;
\item b) a small ``boundary layer''  of width $\eta^{1/3}$ surrounding the 
band; the trading speed is on the order $\eta^{-1/3}$ and takes a scaling form;
\item c) further away from the band, but still within its zone of influence, the trading speed is on the order $\eta^{-1/2}$ and behaves as a square-root of the distance to the band; 
\item d) finally, far away from the band, the trading speed is a linear function of the distance to the ideal position and one recovers the exact $\Gamma=0$ solution as expected. 
\end{itemize}
Our method in fact readily generalizes to other non-linear cost structures and in particular to the $Q^{3/2}$ law alluded to above. We briefly discuss how our results extend to this case in the 
final section of this paper.

\section{Set up of the problem and the $\Gamma=0$ solution}

Following \cite{MarSch} we assume that the value $X_t$ of the traded instrument has a mean-reverting dynamics governed by the following drift-diffusion equation\footnote{The diffusion constant $\sigma^2$ can also
depend on $X_t$, as in \cite{MarSch}, without materially impacting the following results. For the sake of simplicity, we keep $\sigma$ constant.}
\be
dX_t = \mu(X_t) {\rm d}t + \sigma {\rm d}W_t\,. 
\ee
We will call $\mathcal{L}_x$ the associated It\^{o} differential operator
\beq
\mathcal{L}_x[f] = \mu(x) \frac{\partial f}{\partial x} + \frac{\sigma^2}{2} \frac{\partial^2 f}{\partial x^2}\,.
\eeq
The position (number of shares/lots, etc.) of the manager at time $t$ is denoted by $\theta_t$. For a given rebalancing policy, the expected risk-adjusted P\&L per unit time, conditional on $X_t=x$ and $\theta_t=\theta$, is given by
\be
W(\theta,x|t) = \lim_{T \to \infty} \frac1T \int_t^{\infty} {\rm d}t' \, e^{-(t'-t)/T} \, \left[ \mu(X_{t'})\,\theta_{t'} - \lambda\,\theta^2_{t'} - \Gamma\,|\dot{\theta}_{t'}|-\eta\, \dot{\theta}^2_{t'} \right]\,,
\ee
where $\lambda$ is the cost of risk (that includes a factor $\sigma^2$). The first term is the average gain of the position, the last two terms are rebalancing costs. We now introduce the value function $V(\theta,x)$, defined
as $W(\theta,x|t)$ for the optimal future rebalancing policy. Note that because we assume a stationary process for $X_t$, 
the value function is in fact independent of $t$ \footnote{Technically, it can be useful to keep $T$ large but finite, which amounts to regularizing the differential operator 
$\mathcal{L}_x[f]$ with a term $- f/T$ and taking the limit $T \to \infty$.}. As is well know, $V(\theta,x)$
then obeys an HJB equation that in the present case reads
\beq \label{HJB}
0 = \mu(x)\,\theta - \lambda\,\theta^2 + \max_{\dot{\theta}} \Big\{  -\Gamma\,|\dot{\theta}|-\eta\, \dot{\theta}^2 +\frac{\partial V}{\partial \theta} \dot{\theta}+ \mathcal{L}_x[V] \Big\}\,.
\eeq
The maximisation with respect to $\dot{\theta}$ is very simple and leads to:
\be
v= \dot{\theta}^* = \frac{1}{2 \eta} \left[\frac{\partial V}{\partial \theta} - \Gamma \text{sign}(\dot{\theta}^*)\right] \quad {\text{or}} \quad v = 0\,,
\ee
where the NT region ($v = 0$) is defined by $|\frac{\partial V}{\partial \theta}| \leq \Gamma$. In this region the HJB equation simplifies to
\beq \label{HJBNT}
\mathcal{L}_x[V_{\text{NT}}]  = - \mu(x)\,\theta + \lambda\,\theta^2 \,.
\eeq
In the rebalancing (RB) region, on the other hand, the HJB equation becomes a non-linear PDE equation
\beq \label{HJBDT}
\left(\frac{\partial V^\pm}{\partial \theta} \pm \Gamma \right)^2= 4 \eta^2 v^2 = 
4 \eta \left[\lambda\theta^2 - \mu(x) \theta   - \mathcal{L}_x[V^\pm]\right]\,,
\eeq
where the $\pm$ sign corresponds, respectively, to large enough positive $\theta$s such that the optimal policy is to sell ($v < 0$), or to large enough negative 
 $\theta$s such that the optimal policy is to buy ($v > 0$). 

\section{The $\eta = 0$ solution}

For $\eta = 0$, the solution of the corresponding HJB equation has been worked out by Martin \& Sch\"{o}neborn (M\&S) in \cite{MarSch}, and will be denoted by $V_0(\theta,x)$. The NT region is 
parameterized by two functions $\Theta^-_0(x),\Theta^+_0(x)$, such that for a given $x$, the speed of trading $v$ vanishes inside the interval $[-\Theta^-_0(x),\Theta^+_0(x)]$, hereafter referred to as a band. Outside the 
band, the $\eta=0$ solution to Eq. (\ref{HJBDT}) is given by
\beq \label{sol0DT}
V_0^\pm(\theta,x) = V_{\text{NT},0}(\pm \Theta^\pm_0(x),x) - \Gamma \left|\theta  \mp \Theta^\pm_0(x)\right|\,.
\eeq
Inside the band, the general solution to the linear equation (\ref{HJBNT}) can be constructed using the Green's function $\mathcal{G}_x$ of the operator $\mathcal{L}_x$ and the two independent solutions $\psi_{1,2}(x)$
of the homogeneous equation $\mathcal{L}_x f = 0$, see \cite{MarSch} for details. Schematically,
\be\label{gal-solution}
V_{\text{NT},0}(\theta,x) = \mathcal{G}_x[ - \mu(x)\,\theta + \lambda\,\theta^2] + \alpha_1(\theta) \psi_1(x) + \alpha_2(\theta) \psi_2(x)\,,
\ee
where $\alpha_{1,2}(\theta)$ are two yet-to-be-determined functions. The reference \cite{MarSch} proposes to fix these functions in two steps. First, one imposes that at the (still unknown) boundaries of the NT zone, 
the derivative of $V_0(\theta,x)$ are continuous, i.e.
\be
\left.\frac{\partial V_{\text{NT},0}(\theta,x)}{\partial \theta}\right|_{\theta = \pm \Theta^\pm_0(x)} = \mp \Gamma\,.
\ee
This allows one to solve for $\alpha_{1,2}(\theta)$ as functionals of the boundary positions $\Theta^\pm_0(x)$. Second, one determines these boundaries by invoking the variational argument, i.e that 
these boundaries should maximize the value function $V_{0,\text{NT}}(\theta,x)$ everywhere in the NT region. This second condition allows to fully determine $\Theta^\pm_0(x)$. While we fully
agree with the final expressions obtained by M\&S, we argue that their second condition does not generalize to the case $\eta \neq 0$. The general condition should rather be that the second 
derivative of the value function with respect to $\theta$ is continuous everywhere, including the boundaries between NT and RB regions -- see Appendix A. In fact, we show in Appendix B that the M\&S solution obeys
\be
\left.\frac{\partial^2 V_{\text{NT},0}(\theta,x)}{\partial \theta^2}\right|_{\theta = \pm \Theta^\pm_0(x)} \equiv 0\,,
\ee
a property that apparently went unnoticed in \cite{MarSch} and that is actually much simpler than the variational condition. 
In the next section we will attempt to construct a consistent solution to the HJB for arbitrary $\Gamma$ 
but small quadratic costs $\eta \to 0$. We will make use of the continuity of the first and second derivative of the value function to determine the new location of the boundaries.

\section{The small $\eta$ matched asymptotic expansions}

\subsection{The outer region}

Let us assume that, far enough from the new band positions $[-\Theta^-_\eta(x),\Theta^+_\eta(x)]$ (called the ``outer region''), the trading solution for small $\eta$ reads:
\beq
V_\eta^\pm(\theta,x) = V_0^\pm(\theta,x) + \sqrt{\eta} V_1^\pm(\theta,x) + \dots \qquad (\eta \to 0)\,.
\eeq
In the sequel we will confine ourselves to the `$+$' sector where the trading speed is negative and we will drop the superscripts on the position of the
band $\Theta_0$ and on $V_0,V_1$. Plugging our ansatz into Eq. (\ref{HJBDT}) and retaining leading terms in $\eta$ gives
\be
\frac{\partial V_1}{\partial \theta} = - 2 \sqrt{\lambda\theta^2 - \mu(x) \theta - \mathcal{L}_x[V_0]}\,,
\ee
where we have used that $\frac{\partial V_0}{\partial \theta} = -\Gamma$ so that the zeroth-order term in the LHS vanishes. Using the 
solution outside the band \eqref{sol0DT} we write
\be
\mathcal{L}_x[V_0] = \mathcal{L}_x[V_{\text{NT},0}(\Theta_0(x),x)] + \Gamma \mathcal{L}_x[\Theta_0(x)] \,.
\ee
Note that
\be
\mathcal{L}_x[V_{\text{NT},0}(\theta,x)]  = - \mu(x)\,\theta + \lambda\,\theta^2\,,
\ee
from which we deduce (the dependence of $\mu$ and $\Theta_0$ on $x$ is henceforth suppressed)
\beqa 
\mathcal{L}_x[V_{\text{NT},0}(\Theta_0, x)] &=&  -\mu \Theta_0 + \lambda \Theta_0^{2} + \mu \Theta_0' (V_{\text{NT},0})_\theta 
\\ \nonumber
&+&
\frac{\sigma^2}{2} \left[\Theta_0'^2 (V_{\text{NT},0})_{\theta\theta} + \Theta_0'' (V_{\text{NT},0})_{\theta} + 2 \Theta_0' (V_{\text{NT},0})_{x\theta} \right]\,,
\eeqa
where the ``prime'' stands for derivatives wrt $x$ and the subscripts indicate variables (other than $x$) with regard to which derivatives are taken. The boundary conditions at $\Theta_0$ are
\be
\qquad \forall x : \qquad (V_{\text{NT},0})_{\theta}(\Theta_0(x),x) \equiv - \Gamma, \qquad (V_{\text{NT},0})_{\theta \theta}(\Theta_0(x),x) \equiv 0\,, 
\ee
implying
\be
(V_{\text{NT},0})_{x\theta} = - \Theta_0' (V_{\text{NT},0})_{\theta \theta} \equiv 0\,,
\ee
so that $\mathcal{L}_x[V_0]$ finally simplifies to:
\be
\mathcal{L}_x[V_0] \equiv  -\mu \Theta_0 + \lambda \Theta_0^{2}\,.
\ee
Therefore, the equation for $V_1$ becomes:
\be
\frac{\partial V_1}{\partial \theta} =  - 2 \sqrt{\lambda(\theta^2 - \Theta_0^{2}) - \mu (\theta - \Theta_0)}\,.
\ee
Now, the velocity in the trading zone is simply
\be
v = \frac{1}{2\sqrt{\eta}} \frac{\partial V_1}{\partial \theta}\,,
\ee
We thus find that: a) $v$ diverges as $\eta^{-1/2}$ when $\eta \to 0$ recovering the instantaneous rebalancing in this limit; b) $v$ behaves linearly for large $|\theta|$; 
c) $v$ behaves as a square-root close to (see discussion below) the unperturbed band $\Theta_0$
\be
v \approx - \sqrt{\frac{2 \lambda \Theta_0 - \mu}{\eta}}\,\, \sqrt{\theta - \Theta_0}\,.
\ee
This square root singularity is interesting because it means that there must be a region very close to the band where this naive 
perturbative solution breaks down. Indeed, the second derivative of $\sqrt{\eta} V_1$ wrt $\theta$ diverges and thus may not be neglected. One has to analyse this region by zooming in on the immediate
proximity of the band conventionally dubbed the ``inner region'' or the boundary layer, see \cite{Hinch}.

\subsection{The inner region}

To make a start on the analysis, we take the derivative of Eq. (\ref{HJBDT}) with respect to $\theta$ and introduce $f := 
\partial V^+/\partial \theta + \Gamma \equiv 2 \eta v$. This leads to the exact equation
\be\label{HJBprime}
f f_\theta = -2 \eta (\mu - 2 \lambda \theta + \mathcal{L}_x[f])\,.
\ee
We postulate that close to the (new) band $\Theta_\eta(x)$ the function $f$ exhibits the following scaling
\be
f = - \eta^\alpha F(y)\,, \qquad \textrm{where} \qquad y :=  \frac{\theta - \Theta_\eta(x)}{\eta^\beta} > 0, \qquad \alpha, \beta > 0\,.
\ee
The parameters $\alpha$ and $\beta$ are two exponents that need to be determined and $F(y)$ is a positive function (note indeed that $f \propto v < 0$ in the `+' sector that
we are considering here). A first condition comes from the fact that when $y \to \infty$, 
this scaling form must reproduce the above square-root solution, i.e., $F(y) \simeq A \sqrt{y}$ for large $y$. This requires
\be
\eta^{\alpha - \beta/2} A = 2\sqrt{\eta} \sqrt{2 \lambda \Theta_0 - \mu}\,,
\ee
leading to
\be
\alpha - \beta/2 = 1/2\,, \qquad A = 2 \sqrt{2 \lambda \Theta_0 - \mu}\,.
\ee
Injecting the scaling form into Eq. (\ref{HJBprime}) and noting that derivatives with respect to $x$ supply factors of $\eta^{-\beta}$, 
we find that the {\it leading} terms are
\be
\eta^{2 \alpha - \beta} F F_y = 2 \eta \left(-\mu + 2 \lambda \Theta_0 + \frac{\sigma^2}{2} \eta^{\alpha- 2 \beta} \Theta_0'^2 F_{yy}\right)\,.
\ee
Matching the $\eta$ powers of the two sides of the equation leads to
\be
2 \alpha - \beta = 1 \qquad \text{or} \qquad  \,\, 2 \alpha - \beta = 1 + \alpha - 2 \beta\,.
\ee
The first equality can only hold if $\alpha \geq 2 \beta$, in which case the last term in the RHS is negligible. This would however lead to
\be
(F^2)_y = 4 (2 \lambda \Theta_0 - \mu) \equiv A^2\,,
\ee
which still has a square-root singularity where $F$ goes to zero, so that close enough to the singularity the last term of the RHS diverges and
cannot be neglected contradicting our original assumption.

The only other possibility is $2 \alpha -\beta =1 + \alpha - 2 \beta$, which, together with $\alpha - \beta/2 = 1/2$, leads 
to $\alpha = 2/3$ and $\beta=1/3$. The leading ODE for $F$ takes the following form
\be
(F^2)_y - B F_{yy} = A^2\,, \qquad B \equiv 2\sigma^2 \Theta_0'^2\,.
\ee
Upon integration
\be
F^2 - B F_y = A^2 (y - y_0)\,.
\ee
The solution to the above may be expressed in terms of Airy functions. Writing $F = - B \Psi_y/\Psi$, the above equation reads
\be
\Psi_{yy} = \frac{A^2}{B^2} (y-y_0) \Psi\,.
\ee
We want a solution to this equation such that $\Psi_y(y=0)=0$ (i.e. $F(y=0)=0$) such that $F(y>0) > 0$. The general solution is
\be
\Psi(y) = c_1 \text{Ai}(z (y-y_0)) + c_2 \text{Bi}(z (y-y_0)), \qquad z = \left(\frac{A}{B}\right)^{2/3}\,,
\ee
but the asymptotic behaviour of the Airy functions imposes $c_2 = 0$. The condition $\text{Ai}'(-z y_0) = 0$ selects the first maximum of 
$\text{Ai}$ that occurs for $-zy_0 \approx -1.018..$, thereby fixing $y_0$. Finally, the sought-after solution is
\be\label{sol_scal}
F(y) = - B z \frac{\text{Ai}'(z (y-y_0))}{\text{Ai}(z (y-y_0))}\,.
\ee
For large $y$, one uses the asymptotic behaviour $\ln \text{Ai}(u) \approx -2u^{3/2}/3$ to obtain
\be
F(y) \approx  B z^{3/2} \sqrt{y} \equiv A \sqrt{y}\,,
\ee
as desired. We have found a solution that goes smoothly to zero when $y \to 0$, i.e. in a region of width $\eta^{1/3}$ 
immediately outside the boundary of the band $-\Theta_\eta$, see Figure \ref{fig:solution}. Note that $F_y(0)= 1.018..A^{4/3} B^{-1/3}$.  

Note that in the small $\Gamma$ limit, it is well know that the no-trade region has a width of order $\Gamma^{1/3}$ around the ideal (Markowitz) position, therefore leading to
$2 \lambda \Theta_0 - \mu \sim \Gamma^{1/3}$, or $A \sim \Gamma^{1/6}$ and $z \sim \Gamma^{1/9}$. Plugging this into Eq. (\ref{sol_scal}) leads to a width of the inner region 
scaling as $(\eta^3/\Gamma)^{1/9}$. This allows us to make the connection between our results and those of Ref.  \cite{LiuKarWeb}, where the authors consider the double limit $\Gamma, \eta \to 0$
with ${\eta}/{\Gamma^{4/3}}$ fixed. In that regime, $(\eta^3/\Gamma)^{1/9} \sim \Gamma^{1/3}$, indeed recovering the predictions of  \cite{LiuKarWeb}. Our above results are, we believe, quite interesting
as they are valid for arbitrary values of $\Gamma$, with a universal shape of the scaling function $F(y)$ in the inner region, independent of the precise problem at hand.  

\subsection{Boundaries shift inwards}

Next, we need to find the shifted band position $\Theta_\eta$. We will use the fact that the second derivative of the value function should be continuous at the boundary.
Integrating $f$ leads to the following equality for the value function in the trading zone
\be\label{new-sol}
V^+(\theta,x) = \int_{\Theta_\eta}^\theta {\rm d}\theta' f(\theta',x) - \Gamma (\theta - \Theta_\eta) + V_{\text{NT},\eta}(\Theta_\eta,x)\,,
\ee
that by construction coincides with $V_{\text{NT},\eta}$ at the boundaries of the band. Observe that $V_{\text{NT},\eta}$ differs from the original M\&S solution by the change of boundary conditions.  
Because of the optimality of the $\eta=0$ boundaries, one immediately infers that $|\Theta_\eta - \Theta_0| = O(\eta^a)$ translates to $|V_{\text{NT},\eta}(\theta)-V_{\text{NT},0}(\theta)| = O(\eta^{2a})$. 

We will now show that $a=1/3$, i.e. that the shift of the band is of the same order as the width of the boundary layer. 
The continuity of the second derivative at $\theta = \Theta_\eta$ imposes, to leading order
\be \label{balance}
- \eta^{1/3} F_y(0) = (V_{\text{NT},0})_{\theta \theta}(\Theta_\eta) \approx (V_{\text{NT},0})_{\theta \theta \theta}(\Theta_0) (\Theta_\eta - \Theta_0)\,,
\ee
where we used the condition $(V_{\text{NT},0})_{\theta \theta}(\Theta_0) = 0$ derived in Appendix B. This immediately yields
\be
\Theta_\eta = \Theta_0 - \frac{F_y(0)}{(V_{\text{NT},0})_{\theta \theta \theta}(\Theta_0)} \, \eta^{1/3}\,,
\ee
i.e. the inward shift of the band is on the order $\eta^{1/3}$ exactly as the width of the boundary layer. This is reasonable as it means that both effects of the quadratic term on the immediate proximity of the band are of the same order. As we show in Appendix B, the third derivative of $V_{\text{NT},0}$
turns out to be positive at the unperturbed band, hence the above expression implies that the boundary shifts inwards, i.e. the NT region is {\it reduced} by the presence
of quadratic costs. 

The alert reader might wonder how the solution given by Eq. (\ref{new-sol}) above still has a continuous first derivative at $\Theta_\eta$. In fact, from the 
vanishing of the second derivative of the M\&S solution at the unperturbed band, the Taylor expansion gives: 
\be
(V_{\text{NT},0})_{\theta}(\Theta_\eta) \approx (V_{\text{NT},0})_{\theta}(\Theta_0) + 0 + \frac12 (V_{\text{NT},0})_{\theta \theta \theta}(-\Theta_0) (\Theta_0 - \Theta_\eta)^2\,,
\ee
with $(V_{\text{NT},0})_{\theta}(\Theta_0) \equiv -\Gamma$. Now, since the HJB equation itself is independent of $\eta$ in the NT region, its solution can only 
depend on $\eta$ through boundary conditions. But because of the optimality of the position of the $\eta=0$ boundary, we have everywhere inside the band
\be
V_{\text{NT},\eta}(\theta) = V_{\text{NT},0}(\theta) + \frac12 (\Theta_0 - \Theta_\eta)^2 g(\theta)+\ldots\,.
\ee
Here $g(\theta)$ is the second derivative of the M\&S solution with respect to the position of the boundary. 

In view of (\ref{new-sol}) the continuity of the first derivative of the perturbed solution across the band is guaranteed by
\be
-\Gamma \equiv (V_{\text{NT},\eta})_\theta(\Theta_\eta)\,, 
\ee
or, using the two previous equations,
\beqa \nn
-\Gamma &=&(V_{\text{NT},0})_\theta(\Theta_\eta) + \frac12 (\Theta_0 - \Theta_\eta)^2 g'(\Theta_0)+\ldots\\ \nn
&=& (V_{\text{NT},0})_{\theta}(\Theta_0) + \frac12 (\Theta_0 - \Theta_\eta)^2 \left[ (V_{\text{NT},0})_{\theta \theta \theta}(\Theta_0) + g'(\Theta_0)
\right] + \ldots\,.
\eeqa
However, remembering that $(V_{\text{NT},0})_{\theta}(\Theta_0)=-\Gamma$, one finally arrives at the following consistency condition
\be \label{gprime}
g'(\Theta_0) = -(V_{\text{NT},0})_{\theta \theta \theta}(\Theta_0)\,.
\ee
As we show in Appendix B this is indeed a property of the M\&S solution. This establishes that our boundary layer solution (\ref{new-sol}) is $C^2$ across the NT-RB boundary.

\section{Discussion \& extensions}

All the results above are compatible with the numerical results of Ref. \cite{LiuKarWeb}, which exhibit a square-root-like trading speed close to the band and a band that 
shrinks when $\eta$ increases (\textit{c.f.} Figure 1 therein; the parameters $\lambda, \epsilon$ are to be identified with our $\eta, \Gamma$, respectively). We have ourselves solved the HJB equation numerically and 
found a band position compatible with our prediction above $\Theta_0 - \Theta_\eta \propto \eta^{1/3}$. 

In the above we did not elaborate on the domain of validity of the small $\eta$ expansion. For perturbation theory to make sense, the shift in the position of the band must remain small compared to the width of the band itself, i.e.:
\be
\left| \frac{\Theta_0^+ - \Theta_\eta^+}{\Theta_0^+ + \Theta_0^-} \right| \ll 1\,.
\ee
Assuming an Ornstein-Uhlenbeck process for the prediction\footnote{$\Omega^{-1}$ is the mean-reversion time}, $\mu(x) = - \Omega x$, we obtain in the small $\Gamma$ 
limit
\be
(V_{\text{NT},0})_{\theta \theta \theta}(\Theta_0) \approx \frac{8 \lambda^2}{\sigma^2 \Omega} \left(\frac{3 \Gamma \sigma^2}{ 2 \Omega}\right)^{1/3}\,.
\ee
This, together with the results of \cite{MarSch}
\be
\Theta_0^+ + \Theta_0^- \sim \frac{\sigma^2}{\lambda} \left(\frac{\Gamma \Omega^2}{\sigma^4}\right)^{1/3}\,, \qquad \Theta_0' \sim \frac{\Omega}{\lambda}\,,
\ee
where we have dropped all $O(1)$ numerical constants, allows us to recast the above as\footnote{In the following discussion we assume that $\lambda \Theta_0 \ll \mu$, i.e. that 
the price of risk for a position at the edge of the band is small compared to the expected gains.}
\be
\frac{\eta}{\Gamma^{4/3}} \ll  \frac{\lambda}{\left(\sigma \Omega\right)^{4/3}}\,,
\ee
so that the relevant combination is indeed ${\eta}\, {\Gamma^{-4/3}}$, as anticipated in \cite{LiuKarWeb} in the $\Gamma, \eta \to 0$ limit.  In order to make sense of the above inequality, it is useful to substitute the price of risk $\lambda$ with a risk target ${\cal R}$ corresponding to
the volatility of the ideal position. In the absence of costs this ideal position reads
\be
\theta^* = \frac{\mu}{2 \lambda}\,,
\ee
leading to a typical risk ${\cal R} \sim \sqrt{\Omega} \sigma^2/2\lambda$. The above inequality can thus be rewritten as
\be
\frac{\eta}{\Gamma^{4/3}} \ll  \frac{\sigma^{2/3}}{{\cal R} \Omega^{5/6}}\,.
\ee
Suppose that the target risk ${\cal R}$ is a fraction $\varphi$ of the daily volume $V$, i.e. ${\cal R} = \varphi V \sigma$. The quadratic cost parameter may be expressed using dimensionful quantities $\eta = \gamma \sigma T^{3/2}/V$, 
where $\gamma$ is a number and $T=1$ day. The final dimensionless condition is, interestingly enough, independent of the volume $V$
\be
\gamma \varphi \ll \left(\frac{\Gamma}{\sigma \sqrt{T}}\right)^{4/3} (\Omega T)^{-5/6}\,.
\ee
Taking for example $\Gamma = 2$ bp, $\sigma \sqrt{T}= 2 \%$ and $\Omega T=0.1$, we find $\gamma \varphi \ll 10^{-2}$. 
In conclusion, our small $\eta$ expansion makes sense, for a given quadratic cost coefficient $\gamma$, if the portfolio's typical positions represent a small fraction 
of the daily traded volume. 

Let us conclude by pointing out that the above matched asymptotic expansion around the M\&S solution can be undertaken for arbitrary non-linear costs. If instead of the quadratic cost term 
$\eta\, \dot{\theta}^2$ considered above one worked with the more realistic $\zeta\, \dot{\theta}^{3/2}$ term, one would find the following HJB equation in the trading region
\beq \label{HJBcube}
\left(\frac{\partial V^\pm}{\partial \theta} \pm \Gamma \right)^3 \propto \zeta^2 \left[\lambda\theta^2 - \mu(x) \theta   - \mathcal{L}_x[V^\pm]\right]\,,
\eeq
requiring a perturbative expansion of the form
\beq
V_\zeta^\pm(\theta,x) = V_0^\pm(\theta,x) + \zeta^{2/3} V_1^\pm(\theta,x) + \dots \,.
\eeq
This leads to trading speed behaving as $(\theta - \Theta_\zeta)^{1/3}$ in the outer region close enough to the band, but crossing over to the following boundary layer solution
\be
f(\theta) = \zeta^{4/5} G\left(\frac{\theta - \Theta_\zeta(x)}{\zeta^{2/5}}\right)\,,
\ee
where $G(y)$ now obeys the following Abel equation (see e.g. \cite{Abel})
\be
G^3 - B' G_y = A'^2 (y - y_0)\,.
\ee
One should impose $G(y=0)=0$ and $G(y) \sim y^{1/3}$ for large $y$. Since our asymptotic expansion in the quadratic case relies chiefly on the properties of the M\&S solution in the NT 
region, all the results obtained readily transpose to the present case as well.

\subsection*{Acknowledgements} We thank J. Donier, C. A. Lehalle, J. Muhle-Karbe and M. Potters for enlightening discussions on these subjects. 

\section*{Appendix A: Continuity of the second derivative of the value function}

In this appendix we wish to show that the rebalancing (RB) and no-trade (NT) value functions match smoothly at the boundary of the RB-NT regions up to and including second order derivatives. We will suppose that this boundary is smooth, that the value function is continuous and differentiable in each region.
We rewrite the HJB equation as
$$\mathcal{L}_x[V] = -{\eta} v^2(x,\theta)+H(x, \theta)\,,$$
with the velocity $v(x,\theta)$ given by
$$
v(x,\theta)=\begin{cases}
(\frac{\partial V}{\partial \theta} + \Gamma)/2\eta & \text{inside the RB zone}\,, \\
0  &\text{inside the NT zone}\,.
\end{cases}
$$
We write the remainder of the right hand side as  $H(x, \theta)$ as it is a regular function and does not play any role in matching the two regions.

Along the $x$ direction $v(x, \theta)$ is at most discontinuous and, since $\mathcal{L}_x$ is a second-order linear operator, this implies that $\frac{\partial V}{\partial x}$ is continuous at the boundary. As we suppose that $V$ is continuous across the boundary, this implies that its derivatives across the boundary are continuous if we assume that the boundary is not collinear with $x$. Therefore $v(x,\theta)$ must be continuous.
By the same token, the continuity of $v(x,\theta)$ implies continuity of $\frac{\partial^2 V}{\partial x^2}$.

Differentiating the HJB equation with respect to $\theta$ gives
$$\mathcal{L}_x[V_\theta] = -{2 \eta} v_\theta(x,\theta) v(x,\theta)+H_\theta(x, \theta)\,.$$
In the right hand side, $ v_\theta(x,\theta)$ is discontinuous when crossing the boundary, but the product $v_\theta(x,\theta) v(x,\theta)$ remains continuous since $v(x,\theta)$ passes continuously through zero at the boundary.
This implies that the mixed derivative  $\frac{\partial^2 V}{\partial x \partial \theta}$ is continuous, otherwise the term $\frac{\partial^3 V}{\partial x^2 \partial \theta}$ in the left hand side would have to contain a delta function.
Continuity of $V$, its first derivatives and that of $\partial_x \partial_\theta V$ guarantees continuity of $\partial^2_{\theta} V$.

\section*{Appendix B: Some properties of the M\&S solution}

\subsection*{The M\&S solution has a vanishing second derivative at the boundary}

We would like to compute the second derivative of the M\&S solution at the optimal boundary
\beq
\frac{\partial^2 V_\textrm{NT,0}}{\partial \theta^2} \bigg|_{x=\hat{h}_\pm}\,,
\eeq
where $\hat{h}_\pm$ is such that $\Theta_0^\pm(\hat{h}_\pm)=\theta$\,. 

To set the notation we use the decomposition \eqref{gal-solution}
\beq \label{decomp_with_i}
\partial_\theta V_\textrm{NT,0} = I(x,\theta)  + \alpha'_1(\theta) \psi_1(x) + \alpha'_2(\theta) \psi_2(x)\,.
\eeq
The first term represents the theta derivative of the particular solution
\beq
I(x,\theta) = \int^\infty_{-\infty} (\mu(\xi) - 2\,\lambda \,\theta) {\cal G}(x,\xi) d\xi\,.
\eeq
The derivatives functions $\alpha'_1, \alpha'_2$ are given by (see \cite{MarSch})
\beqa \nn
D \alpha'_1(\theta) &=&  -\Gamma \psi_2(h_-) - \Gamma \psi_2(h_+) + I(h_-,\theta) \psi_2(h_+) - I(h_+,\theta) \psi_2(h_-)\,,  \\
\label{alpha1} \\ \nn
D \alpha'_2(\theta) &=&  \Gamma \psi_1(h_-) + \Gamma \psi_1(h_+) + I(h_+,\theta) \psi_1(h_-) - I(h_-,\theta) \psi_1(h_+)\,,  \\
\label{alpha2} 
\eeqa
where the determinant $D \equiv D(h_+,h_-) = \psi_1(h_+)\psi_2(h_-) - \psi_1(h_-)\psi_2(h_+)$. The optimal boundaries are defined by the variational equality
\beq
\frac{\partial \alpha'_{1,2} }{\partial h_\pm} \Bigg|_{h_\pm = \hat{h}_\pm} = 0\,.
\eeq
Consequently,
\beq \label{d_of_aprime}
\frac{d \alpha'_{1,2} } {d \theta} \Bigg|_{h_\pm = \hat{h}_\pm} =  \frac{\partial \alpha'_{1,2} }{\partial \theta} \Bigg|_{h_\pm = \hat{h}_\pm}\,. 
\eeq
We now have all the ingredients to compute the second derivative at the optimal boundaries
\beq
\partial^2_\theta V_\textrm{NT,0}(\hat{h}_+,\theta) = \partial_\theta I(\hat{h}_+,\theta) + \frac{d \alpha'_{1} } {d \theta} \Bigg|_{h_\pm = \hat{h}_\pm} \psi_1(\hat{h}_+) + \frac{d \alpha'_{2} } {d \theta} \Bigg|_{h_\pm = \hat{h}_\pm} \psi_2(\hat{h}_+)\,.
\eeq
Plugging the formulae \eqref{alpha1}-\eqref{alpha2} and using \eqref{d_of_aprime} we indeed find:
\beq
\partial^2_\theta V_\textrm{NT,0}(\hat{h}_+,\theta) = 0\,,
\eeq
as expected from the continuity of the second derivative of the value function of the boundary. Remember that the M\&S solution is a linear function of $\theta$ in the trading zone, therefore
$\partial^2_\theta V$ is identically zero there. A similar result may be found at $x=\hat{h}_-$. 

\subsection*{A useful identity}

Consider again the derivative of the M\&S solution at the band \eqref{decomp_with_i}. The derivative $\partial_\theta V_\textrm{NT,0}$ must be equal to $\mp \Gamma$ at $\theta=\pm \Theta_0^\pm$. A change to any of the parameters of the problem will result in the boundaries being shifted by a $\delta \Theta$. This induces a small change of the functions $\alpha_{1,2}$ that we denote $\delta \alpha_{1,2}$. Since $I(x,\theta)$ does not depend explicitly on the boundary and is linear in $\theta$, maintaining the correct boundary condition
to second order in $\delta \Theta$ leads to the following identity
\be
\delta \alpha'_1 \psi_1(x) + \delta \alpha'_2 \psi_2(x) + \frac12 \,(\delta \Theta)^2 \left[\alpha'''_1 \psi_1(x) + \alpha'''_2 \psi_2(x)\right]= 0\,,
\ee
where all quantities are evaluated at one of the boundaries. This is equivalent to \eqref{gprime}.

\subsection*{Sign of the third derivative}

Our strategy will be to take the derivative of the boundary conditions
\beqa
\mp \Gamma &=& I(x, \pm \Theta_0^\pm)  + \alpha'_1(\pm \Theta_0^\pm) \psi_1(x) + \alpha'_2(\pm \Theta_0^\pm) \psi_2(x)\,, \\
0 &=& I'(x,\pm \Theta_0^\pm)  + \alpha''_1(\pm \Theta_0^\pm) \psi_1(x) + \alpha''_2(\pm \Theta_0^\pm) \psi_2(x)\,,
\eeqa
with respect to the linear cost parameter $\Gamma$. We will present the argument for the upper boundary. Introducing
\be
\rho_x = \frac{\partial \Theta_0^+}{\partial x} < 0\,, \qquad \rho_\Gamma = \frac{\partial \Theta_0^+}{\partial \Gamma} > 0\,,
\ee
the derivative of the first equation may be written as
\beqa \nn
1 &=& \rho_\Gamma \left[I'(x,\Theta_0^+) + \alpha''_1(\Theta_0^+) \psi_1(x) + \alpha''_2(\Theta_0^+) \psi_2(x)\right] + \\
&&+
\frac{\partial \alpha'_1}{\partial \Gamma}(\Theta_0^+) \psi_1(x) + \frac{\partial \alpha'_2}{\partial \Gamma}(\Theta_0^+) \psi_2(x)\,.
\eeqa
Note that because of the second boundary condition the term inside the brackets vanishes. Taking the derivative of the resulting equation wrt $x$ now leads to
\beqa
0 &=& \rho_x \left[ \frac{\partial \alpha''_1}{\partial \Gamma}(\Theta_0^+) \psi_1(x) + \frac{\partial \alpha''_2}{\partial \Gamma}(\Theta_0^+) \psi_2(x)
\right] +\\
&+& \frac{\partial \alpha'_1}{\partial \Gamma}(\Theta_0^+) \psi'_1(x) + \frac{\partial \alpha'_2}{\partial \Gamma}(\Theta_0^+) \psi'_2(x) \equiv J_1 + J_2 \,.
\eeqa
One can use the explicit solutions for $\alpha_{1,2}$ in \eqref{alpha1}-\eqref{alpha2} to compute the partial derivatives with respect to $\Gamma$ by observing that the implicit dependence through the boundaries cancels out because of the M\&S optimality condition. We have
\be
\frac{\partial \alpha'_1}{\partial \Gamma} = - \frac1D \left( \psi_2(h_-) + \psi_2(h_+) \right)\,, \quad
\frac{\partial \alpha'_2}{\partial \Gamma} =  \frac1D \left( \psi_1(h_-) + \psi_1(h_+) \right)\,,
\ee
so that the second term $J_2$ in the last equation above reads
\beqa
J_2 &=& - \frac1D \left(\psi_2(h_+) \psi'_1(h_+) - \psi'_2(h_+) \psi_1(h_+) + \right. \\
&&\left. \psi_2(h_-) \psi'_1(h_+) - \psi_1(h_-) \psi'_2(h_+)\right)\,.
\eeqa
Interestingly enough, the first two terms in the parenthesis is the Wronskian which is positive, \textit{c.f.} \cite{MarSch}. The last two terms also give a positive 
contribution since they can be written as $\partial_{h_+} D$ which is positive when $h_+ > h_-$. Consequently, $J_2 < 0$ and 
therefore $J_1 > 0$. Now, take the derivative of the second boundary condition wrt $\Gamma$
\be
0= \rho_\Gamma \left[\alpha'''_1(\Theta_0^+) \psi_1(x) + \alpha'''_2(\Theta_0^+) \psi_2(x)\right] + 
 \frac{\partial \alpha''_1}{\partial \Gamma}(\Theta_0^+) \psi_1(x) + \frac{\partial \alpha''_2}{\partial \Gamma}(\Theta_0^+) \psi_2(x)\,.
\ee
One immediately infers that
\be
\rho_x \rho_\Gamma \left[\alpha'''_1(\Theta_0^+) \psi_1(x) + \alpha'''_2(\Theta_0^+) \psi_2(x)\right] = - J_1\,, 
\ee
which in view of $\rho_x \rho_\Gamma < 0$, leads one to conclude
\be
V_{\theta\theta\theta}(\Theta_0^+) \equiv \alpha'''_1(\Theta_0^+) \psi_1(x) + \alpha'''_2(\Theta_0^+) \psi_2(x) > 0\,.
\ee

\end{document}